\documentclass[aps,prd,eqsecnum,preprint,tightenlines,nofootinbib,showpacs]{revtex4-1}
\usepackage{graphicx,amsmath,latexsym}

\newcommand{\bea}{\begin{eqnarray}}
\newcommand{\eea}{\end{eqnarray}}
\newcommand{\be}{\begin{equation}}
\newcommand{\ee}{\end{equation}}
\newcommand{\ub}[1]{\underline{#1}}
\newcommand{\ob}[1]{\overline{#1}}
\newcommand{\Pminus}{{\cal P}^-}
\newcommand{\veck}{\vec{k}_\perp}
\newcommand{\veckp}{\vec{k}_\perp^{\,\prime}}

\begin{document}

\title{An illustration of the light-front coupled-cluster method
in quantum electrodynamics\footnote{Presented 
at QCD@Work2012, the International Workshop on QCD Theory and Experiment,
June 18-21, 2012, Lecce, Italy.}}

\author{Sophia S. Chabysheva}
\affiliation{Department of Physics \\
University of Minnesota-Duluth \\
Duluth, Minnesota 55812}

\date{\today}

\begin{abstract}
A field-theoretic formulation of the exponential-operator technique
is applied to a nonperturbative Hamiltonian eigenvalue problem in
electrodynamics, quantized in light-front coordinates.  Specifically, we
consider the dressed-electron state, without positron contributions but
with an unlimited number of photons, and compute its anomalous magnetic
moment.  A simple perturbative solution immediately yields the
Schwinger result of $\alpha/2\pi$.  The nonperturbative solution,
which requires numerical techniques, sums a subset of corrections to
all orders in $\alpha$ and incorporates additional physics.
\end{abstract}


\maketitle


\section{Introduction}

The light-front coupled-cluster (LFCC) method~\cite{LFCClett} is
a nonperturbative scheme for the solution of field-theoretic
Hamiltonian eigenvalue problems.  It is based on light-front
quantization~\cite{Dirac,DLCQreviews} and the mathematics of
the many-body coupled-cluster method~\cite{CCorigin} used in
nuclear physics and quantum chemistry~\cite{CCreviews}.  Here
we give a brief introduction to the LFCC method and show how
it can be applied to quantum electrodynamics~\cite{LFCCqed}.

The use of light-front quantization is crucial, in order to
have well-defined Fock-state expansions and separation of
internal and external momenta~\cite{DLCQreviews}.  
We define the coordinates
as $x^\pm=t\pm z$, with $x^+$ the light-front time, and
$\ub{x}=(x^-,x,y)$.  The light-front energy is $p^-=E-p^z$
and the momentum is $\ub{p}=(p^+,\vec p_\perp)$, with
$p^+=E+p^z$ and $\vec p_\perp=(p^x,p^y)$.  The mass-shell
condition $p^2=m^2$ is then $p^-=\frac{m^2+p_\perp^2}{p^+}$,
and the Hamiltonian eigenvalue problem is
\be \label{eq:eigenvalueproblem}
\Pminus|\psi\rangle=\frac{M^2+P_\perp^2}{P^+}|\psi\rangle.
\ee
The eigenstate $|\psi\rangle$ is taken to be an eigenstate
of light-front momentum and expanded in Fock states which
are eigenstates of particle number as well as momentum.
The coefficients of the Fock-state expansion are the wave
functions; these satisfy a coupled set of integral equations
derivable from (\ref{eq:eigenvalueproblem}).

The usual approach to approximation of this infinite
system is to truncate Fock space at some fixed number
of constituents and solve the remaining finite set of
equations.  However, Fock-space truncation can introduce
a number of difficulties~\cite{SecDep}, in particular
uncanceled divergences, which we wish to avoid.  The
LFCC method does avoid them.

The basic idea is to build the eigenstate as
$|\psi\rangle=\sqrt{Z}e^T|\phi\rangle$ from a valence state
$|\phi\rangle$ and an operator $T$.  The $\sqrt{Z}$ factor
maintains the normalization.  The $T$ operator is constructed
to include only terms that increase particle number but
conserve all appropriate quantum numbers, such as momentum
and charge.  
This leads to the definition of an effective Hamiltonian $\ob{\Pminus}=e^{-T}\Pminus e^T$,
which can be constructed from the Baker--Hausdorff expansion
$\ob{\Pminus}=\Pminus+[\Pminus,T]+\frac12 [[\Pminus,T],T]+\cdots$.
The eigenvalue problem can now be written as
\be
P_v\ob{\Pminus}|\phi\rangle=\frac{M^2+P_\perp^2}{P^+}|\phi\rangle, \;\;\;\;
(1-P_v)\ob{\Pminus}|\phi\rangle=0,
\ee
where $P_v$ is a projection onto the valence sector.  The first equation
limits the eigenvalue problem to the valence sector; the second equation
is an equation for the $T$ operator.

At this point, no approximation has been made and the problem remains
of infinite size, because $T$ can contain an infinite number of terms.
The approximation made in the LFCC method is to truncate $T$, rather
than Fock space, and truncate $1-P_v$ to yield a finite set of equations
sufficient to solve for the terms in $T$.

\begin{figure}
\begin{tabular}{c}
\includegraphics[height=1cm]{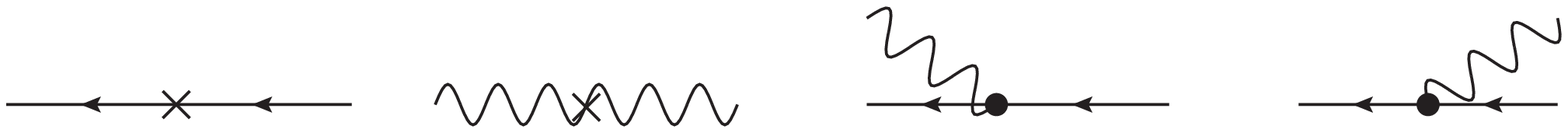} \\
(a) \\
\includegraphics[height=1cm]{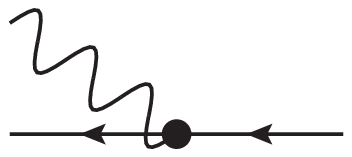} \\
(b) \\
\includegraphics[height=3cm]{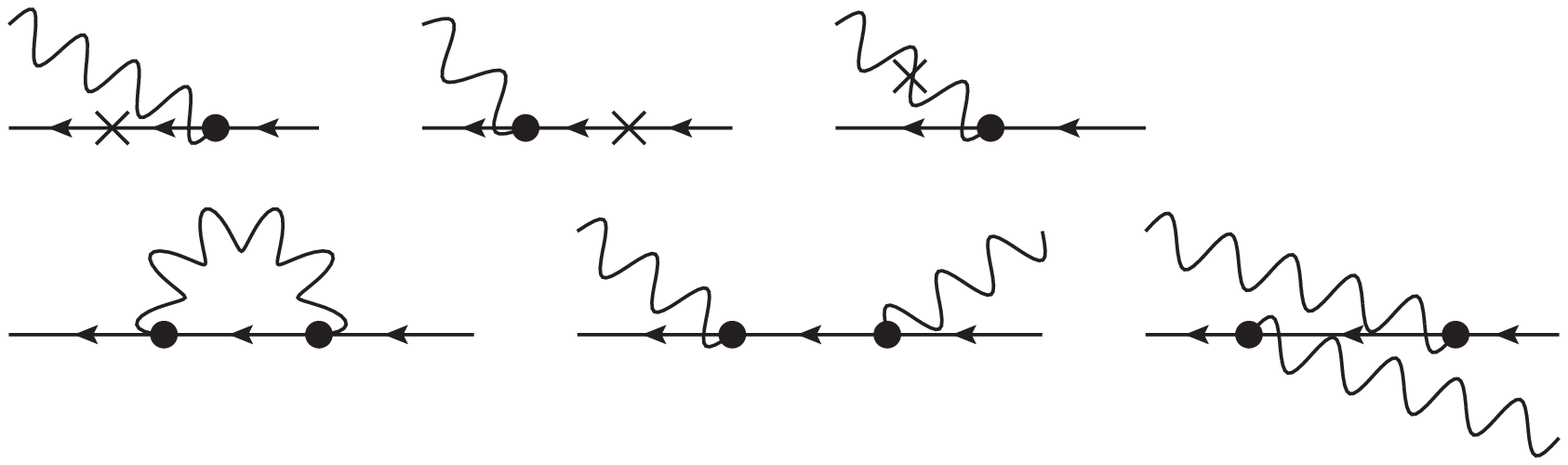} \\
(c) \\
\includegraphics[height=2cm]{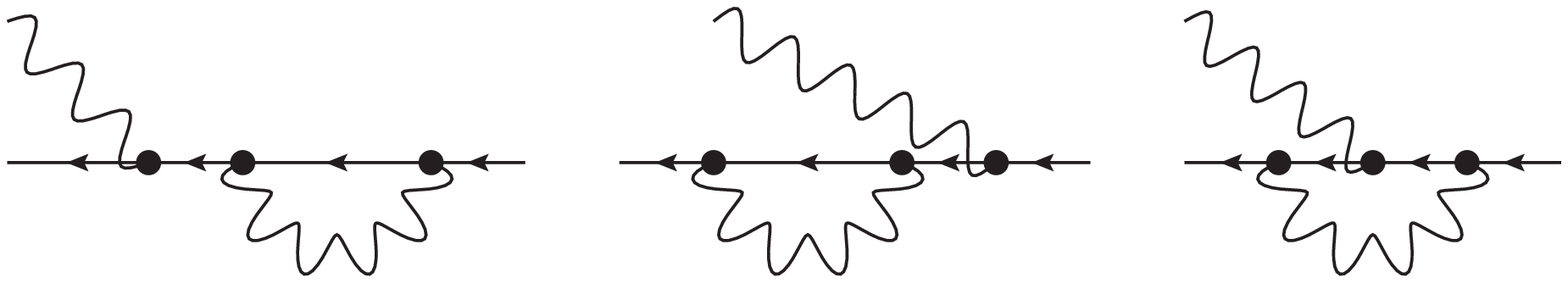} \\
(d)
\end{tabular}
\caption{\label{fig:graphs} Graphs representing (a) the QED Hamiltonian $\Pminus$,
(b) the approximate $T$ operator, and the commutators (c) $[\Pminus,T]$ and 
(d) $[[\Pminus,T],T]$.  Positron contributions are neglected, and
a cross represents a kinetic energy contribution.  In (c) and (d), only terms
that contribute to the given truncation of $1-P_v$ are kept.}
\end{figure}

In the case of the dressed-electron state of QED~\cite{LFCCqed},
the valence state is the bare electron state.
A first choice for $T$ is a term that invokes photon emission,
and the corresponding truncation of $1-P_v$ is to allow only one
additional photon.
If we represent the Hamiltonian $\Pminus$ by the graphs in Fig.~\ref{fig:graphs}(a),
this choice for $T$ can be represented by the graph in Fig.~\ref{fig:graphs}(b).
The commutators that enter the Baker--Hausdorff expansion are represented
in Figs.~\ref{fig:graphs}(c) and (d).  A key result is that the
self-energy loops that appear in Fig.~\ref{fig:graphs} are
all the same, with no spectator or sector dependence.

The truncation of $1-P_v$ limits the number of commutators that need
to be calculated, because each factor of $T$ increases the particle 
number by one.  The QED Hamiltonian includes terms that decrease particle
number by no more than one; therefore, two commutators are sufficient 
when $1-P_v$ limits the increase to one.

\section{Application to QED}

To be more explicit, we give some details of the application of
the LFCC method to QED~\cite{LFCCqed}.  The Pauli--Villars
regulated Lagrangian is~\cite{ArbGauge}
\bea
{\cal L} &=&  \sum_{i=0}^2 (-1)^i \left[-\frac14 F_i^{\mu \nu} F_{i,\mu \nu} 
         +\frac12 \mu_i^2 A_i^\mu A_{i\mu} 
         -\frac12 \zeta \left(\partial^\mu A_{i\mu}\right)^2\right] \\
&&+ \sum_{i=0}^2 (-1)^i \bar{\psi_i} (i \gamma^\mu \partial_\mu - m_i) \psi_i 
  - e \bar{\psi}\gamma^\mu \psi A_\mu, \nonumber
\eea 
with $\psi =  \sum_{i=0}^2 \sqrt{\beta_i}\psi_i$,
  $A_\mu  = \sum_{i=0}^2 \sqrt{\xi_i}A_{i\mu}$, and
  $F_{i\mu \nu} = \partial_\mu A_{i\nu}-\partial_\nu A_{i\mu}$.
Here $i=0$ corresponds to a physical field and $i=1,2$ to Pauli--Villars (PV) fields.
The gauge-fixing parameter $\zeta$ is left arbitrary.
The coupling coefficients are constrained by
$\xi_0=1$,
$\sum_{i=0}^2(-1)^i\xi_i=0$,
$\beta_0=1$,
$\sum_{i=0}^2(-1)^i\beta_i=0$,
and by restoration of chiral symmetry~\cite{ChiralLimit} and a zero
photon mass~\cite{VacPol}.  Neglecting positron contributions,
the Hamiltonian is
\bea \label{eq:QEDP-}
\lefteqn{\Pminus=
   \sum_{is}\int d\ub{p}
      \frac{m_i^2+p_\perp^2}{p^+}(-1)^i
          b_{is}^\dagger(\ub{p}) b_{is}(\ub{p}) 
          +\sum_{l\lambda}\int d\ub{k}
          \frac{\mu_{l\lambda}^2+k_\perp^2}{k^+}(-1)^l\epsilon^\lambda
             a_{l\lambda}^\dagger(\ub{k}) a_{l\lambda}(\ub{k})}&&  \nonumber \\
   && +\sum_{ijl\sigma s\lambda}\int dy d\veck 
   \int\frac{d\ub{p}}{\sqrt{16\pi^3p^+}}     \\
   && \rule{0.3in}{0mm} \times \left\{h_{ijl}^{\sigma s\lambda}(y,\veck)
        a_{l\lambda}^\dagger(yp^+,y\vec{p}_\perp+\veck)
           b_{js}^\dagger((1-y)p^+,(1-y)\vec{p}_\perp-\veck)
                      b_{i\sigma}(\ub{p}) \right. \nonumber \\
&& \left. \rule{0.4in}{0mm}
         +h_{ijl}^{\sigma s\lambda *}(y,\veck)b_{i\sigma}^\dagger(\ub{p})
     b_{js}((1-y)p^+,(1-y)\vec{p}_\perp-\veck)a_{l\lambda}(yp^+,y\vec{p}_\perp+\veck) 
               \right\},  \nonumber
\eea
with $\epsilon^\lambda=(-1,1,1,1)$ and the $h_{ijl}^{\sigma s\lambda}$
known vertex functions~\cite{LFCCqed}.  The $b_{is}^\dagger$ create fermions
of type $i$ and spin $s$, and the $a_{l\lambda}^\dagger$ create photons
of type $l$ and polarization $\lambda$.

The valence state for the dressed electron is just the single-electron state
$|\phi_a^\pm\rangle=\sum_i z_{ai} b_{i\pm}^\dagger(\ub{P})|0\rangle$,
where $a=0,1$; there are two possible states because the valence sector 
includes the PV electron.  The approximate $T$ operator is
\bea
T&=&\sum_{ijls\sigma\lambda}\int dy d\vec{k}_\perp 
   \int\frac{d\ub{p}}{\sqrt{16\pi^3}}\sqrt{p^+} t_{ijl}^{\sigma s\lambda}(y,\vec{k}_\perp) \\
&& \rule{0.5in}{0mm} \times
 a_{l\lambda}^\dagger(yp^+,y\vec{p}_\perp+\vec{k}_\perp)
   b_{js}^\dagger((1-y)p^+,(1-y)\vec{p}_\perp-\vec{k}_\perp)b_{i\sigma}(\ub{p}).
   \nonumber
\eea
The effective Hamiltonian can then be computed~\cite{LFCCqed} and used.

The eigenvalue problem in the valence sector
$P_v \ob{\Pminus}P_v|\phi_a^\pm\rangle=\frac{M_a^2+P_\perp^2}{P^+}|\phi_a^\pm\rangle$
becomes the 2$\times$2 matrix equation
\be
m_i^2 z_{ai}^\pm +\sum_j I_{ij} z_{aj}^\pm = M_a^2 z_{ai}^\pm.
\ee
Here we have the self-energy contribution
\be
I_{ji}=(-1)^i\sum_{i'ls\lambda}(-1)^{i'+l}\epsilon^\lambda
        \int \frac{dy d\veckp}{16\pi^3}
        h_{ji'l}^{\sigma s\lambda*}(y,\vec k_\perp) 
        t_{ii'l}^{\sigma s\lambda}(y,\vec k_\perp).
\ee
The equation for the $T$ operator, obtained from a projection
onto the one-electron/one-photon sector, reduces to
\bea
\lefteqn{\sum_i(-1)^i z_{ai}^\pm\left\{h_{ijl}^{\pm s\lambda}(y,\veck)
+\left[\frac{m_j^2+k_\perp^2}{1-y}+\frac{\mu_{l\lambda}^2+k_\perp^2}{y}-m_i^2\right]
                 t_{ijl}^{\pm s\lambda}(y,\veck) \right.} &&   \\
&& \left. +\frac12 V_{ijl}^{\pm s\lambda}(y,\veck)
+\frac12\sum_{i'} \frac{I_{ji'}}{1-y} t_{ii'l}^{\pm s\lambda}(y,\veck)
-\sum_{j'}(-1)^{i+j'}t_{j'jl}^{\pm s\lambda}(y,\veck)I_{j'i} \right\}=0, \nonumber
\eea
with $V_{ijl}^{\pm s\lambda}$ a vertex correction~\cite{LFCCqed}.

To partially diagonalize in flavor, we define
$C_{abl}^{\pm s\lambda}(y,\veck)
  =\sum_{ij}(-1)^{i+j}z_{ai}^\pm \tilde{z}_{bj}^\pm t_{ijl}^{\pm s\lambda}(y,\veck)$
and use analogous definitions for $H$, $I$, and $V$.  Here the $\tilde{z}_{bj}$
are the left-hand analogs of the $z_{ai}$, which arise because the effective
Hamiltonian is not Hermitian.  The equation
for $C_{abl}^{\pm s\lambda}$ is then
\bea
\lefteqn{\left[M_a^2-\frac{M_b^2+k_\perp^2}{1-y}-\frac{\mu_{l\lambda}^2+k_\perp^2}{y}\right]
   C_{abl}^{\pm s\lambda}(y,\veck)}&& \nonumber \\
&&   =H_{abl}^{\pm s\lambda}(y,\veck)
   +\frac12\left[V_{abl}^{\pm s\lambda}(y,\veck)
      -\sum_{b'}\frac{I_{bb'}}{1-y}C_{ab'l}^{\pm s\lambda}(y,\veck)\right].
\eea
This is to be solved simultaneously with valence sector equations, which depend
on $C/t$ through the self-energy matrix $I$.  Notice that the
physical mass $M_b$ has replaced the bare mass in
the kinetic energy term, without use of sector-dependent renormalization~\cite{SecDep}.

As an example of a calculation of a physical quantity from the dressed-electron
state, we consider the anomalous magnetic moment.  The moment can be obtained
from the spin-flip matrix element of the current
$J^+=\overline{\psi}\gamma^+\psi$ coupled to a photon of momentum $q$.
Computation of such matrix elements requires solution of a left-hand eigenvalue
problem for the effective Hamiltonian; details are given in \cite{LFCCqed}.
When solved perturbatively, this yields the standard Schwinger result~\cite{Schwinger}
of $\alpha/2\pi$ for the anomalous moment.
The full (numerical) solution will include all $\alpha^2$
contributions without electron-positron pairs and partial summation of higher orders.

\section{Summary}

The LFCC method provides a new approach to the nonperturbative 
solution of light-front Hamiltonian eigenvalue problems, one
that avoids Fock-space truncation and its attendant difficulties.
The use of the method is illustrated here in the case of QED~\cite{LFCCqed}
and elsewhere for a soluble model~\cite{LFCClett}.  The approximation
made by truncating the $T$ operator is systematically improvable
by the addition of terms classified by the net number of particles
created and by the total number of annihilation operators.
The self-energy corrections generated in the effective Hamiltonian
are found to be sector and spectator independent.

For QED there are a number of extensions to consider beyond
the present application.  To include positrons, we must first
study the dressed-photon state, in order to set the
photon coupling coefficient at zero photon mass, and then
include pairs in the dressed-electron state.  Once these
eigenstates are computed, we can consider true bound states,
such as muonium and positronium.

Beyond QED, we can, of course, consider QCD.  In fact, nonperturbative
methods are not particularly important for QED and other weak-coupling
theories, and are instead intended for QCD.  There we can begin
with light-front holographic QCD~\cite{hQCD} before considering
the full theory.  It is also interesting to consider simpler theories
with spontaneous symmetry breaking, to better understand the LFCC
method in such a context.


\acknowledgments
This work was done in collaboration with J.R. Hiller
and supported in part by the US Department of Energy.

\end{document}